# Graph-Based Active Machine Learning Method for Diverse and Novel Antimicrobial Peptides Generation and Selection

**Authors**

Bonaventure F. P. Dossou[1,2], Dianbo Liu[1], Xu Ji[1], Moksh Jain[1], Almer M. van der Sloot[3], Roger Palou[3], Michael Tyers[3], Yoshua Bengio[1,4]

**Affiliations**

[1] Mila Quebec AI Institute, 6666 Rue Saint-Urbain, Montréal, QC H2S 3H1, Canada
[2] Jacobs University Bremen, Campus Ring 1, 28759 Bremen, Germany
[3] Institute of Immunology and Cancer, Montréal, QC H3T 1J4, Canada
[4] Canadian Institute for Advanced Research (CIFAR) Senior Fellow

**Abstract**

As antibiotic-resistant bacterial strains are rapidly spreading worldwide, infections caused by these strains are emerging as a global crisis causing the death of millions of people every year. Antimicrobial Peptides (AMPs) are one of the candidates to tackle this problem because of their potential diversity, and ability to favorably modulate the host immune response. However, large-scale screening of new AMP candidates is expensive, time-consuming, and now affordable in developing countries, which need the treatments the most. In this work, we propose a novel active machine learning-based framework that statistically minimizes the number of wet-lab experiments needed to design new AMPs, while ensuring a high diversity and novelty of generated AMPs sequences, in multi-rounds of wet-lab AMP screening settings. Combining recurrent neural network models and a graph-based filter (GraphCC), our proposed approach delivers novel and diverse candidates and demonstrates better performances according to our defined metrics.

**Teaser**

Algorithm for diverse subset selection in the context of active learning for AMP design.

**Introduction**

According to the World Health Organization (WHO) (1), emerging Antimicrobial Resistance (AMR) is causing more than 2.8 million deaths every year and is projected to reach 10 million in 2050. The WHO also alerts that unless immediate actions are taken, we are close to a post-antibiotic era, in which common infections and minor injuries can become again fatal. Therefore, it is paramount to promote research to discover new drugs that would help avoiding a sanitary catastrophe. However, getting a new antimicrobial drug on the market is a delicate, long, and time-consuming process that can cost on average 2.6 billion dollars and ten years of research and development (2). Moreover, because of market failure (3, 4), many large organizations have left the antibiotic discovery space. Therefore, the discovery process requires new approaches to reduce costs and time to discover new antibiotics.

Peptides are short sequences of up to 50-100 amino acids with a variety of biological functions and either have a well-defined structure or, in contrast, can be largely unstructured. Antimicrobial Peptides (AMPs) are found in a large variety of living entities from micro-organisms to humans (5). Some natural AMPs can kill pathogen agents directly, and therefore play key roles in the innate immune system of different organisms (5). Due to the increasing resistance to existing antibiotics, efforts to bring AMPs into clinical use are multiplying (6). Several AMPs are in clinical trials to serve as novel anti-infectives, immunomodulatory agents, and other clinical applications (5, 6).

Machine learning is a sub-field of Artificial Intelligence (AI) that studies and builds systems that learn and improve from experience without being explicitly programmed (7). Active learning (8) (Fig. 1) is a type of learning setup in which a learning algorithm can interactively query an *oracle* to label (annotate, classify) new data points (new generated sequences) with the desired outputs (8).

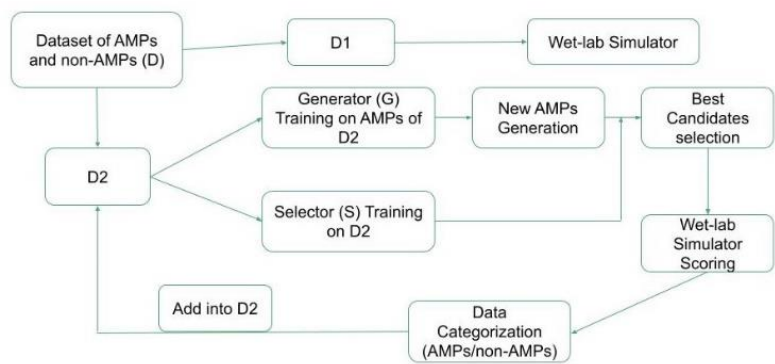

**Fig. 1. Our Active learning framework.** The high-level overview of our active learning framework. An existing dataset of peptide sequences including AMPs and non-AMPs is referred to as "*D*". *D* has been divided into two non-overlapping subsets "*D1*" and "*D2*". A generative model ("*G*") that generates peptide sequences, is trained on AMPs of *D2* and a selector ("*S*") model that predicts $p(y = AMP|x = s)$, defined as the probability of a given peptide sequence *s* to be an AMP. *S* is trained on *D2*. A wet-lab simulator or oracle ("*O*") constructed using AMPs of *D1*, tells how likely a peptide sequence *s* is to be an AMP using its structure and similarity to some pre-defined prototype sequences. Generated peptide structures are filtered by the selection methods baselines (see **Materials and Methods**). Through the selection process, the best sequences are selected and passed to wet-lab simulator *O*. Peptide sequences and their measured effectiveness are added back into *D2* to finetune the models (*G* and *S*).

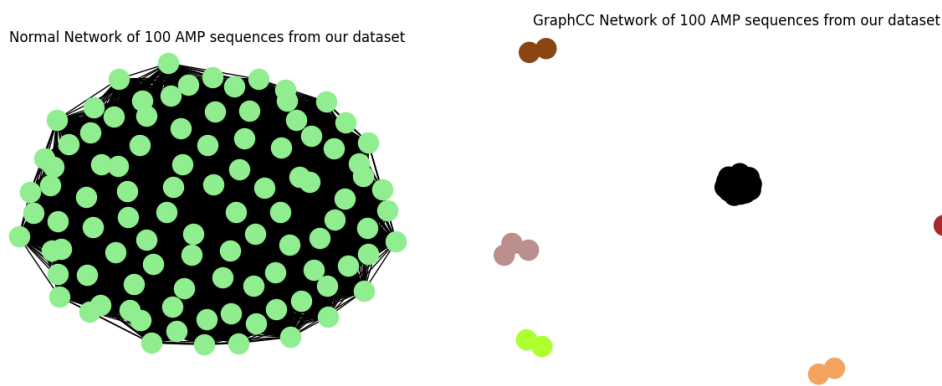

**Fig. 2. a)** Visual description of a normal graph of 100 AMP sequences. **b)** Visual representation of the same graph split in Connected Components (CCs) with 6 CCs: 5 having 3 or less AMPs (long sequences of AMPs), and one CC covering the remaining AMPs (short and middle sequences).

Previous research has demonstrated that active learning can help decide on which examples to label, while improving the performance of the task at hand (8). Deep Learning (DL) based models have gained popularity in recent years and led to the development of systems like AlphaFold2 (9) for protein folding, and ClaSS (10) for accelerated antimicrobial discovery. A DL model trained in an active learning setup can learn from the available data, generalize to unexplored regions of the search space, and build new queries (AMP sequences in our context).

A set of AMP sequences and their similarity relationship can be represented in a graph. A graph is often defined as an ordered pair (*V, E*) where *V* is the set of nodes, and *E* is the set of edges (links – in our context, it represents the edit distance (11) between two sequences) between nodes (Fig. 2a). A graph can be subdivided into distinct, non-overlapping communities of nodes sharing similar properties. Such communities are called *Connected Components (CCs)* (Fig. 2b). In our setting,

AMP sequences are associated with the nodes of our graph, and edges determine whether pairs of AMPs are structurally close to one another, at a distance less than some threshold (see **Materials and Methods**). Based on distances among different AMP molecules, different communities (CCs) are formed. Once such a clustering is performed, representative AMP sequences are sampled from each CC to ensure diversity of the sequences selected to be used as queries sent to the oracle.

In this paper we use Graph Connected Components (*GraphCC*) based active learning to select an experimentally tractable representative set of ~10,000 novel and diverse AMP sequence candidates from a much larger set of generated AMP candidates. This is an important step in drug discovery pipelines where the evaluation of proposed candidates is expensive and needs to be prioritized for the best candidates. Moreover, given high attrition rates typically observed in drug discovery and development, being able to draw from a diverse pool of potential solutions is beneficial. Our approach fosters two important aspects of discovery: diversity and novelty. We describe an active learning framework (Fig. 1) which aims to reduce the number of wet-lab experiments needed, while producing novel and diverse new AMP candidates. This is done through multiple rounds of high throughput wet-lab experiments simulations. Our algorithm can also handle other modular building blocks (like non-natural amino acids etc.) that can be used to directly generate more stable in vivo drugs (peptoids, cyclic peptides, macrocycles). We hope our work allows more efficient/shorter design-test cycles leading to a more rapid generation of potent and pathogen-selective AMPs.

**Results**

**Dataset**

Our dataset of AMP peptides ("*D*") contains amino acid sequences of lengths between 12 and 60, targeting Escherichia Coli, and Candida Albicans bacteria and is obtained from the Database of Antimicrobial Activity and Structure of Peptides (DBAASP) (12). To generate *D*, from the targeted sequences, we considered AMPs (positive sequences) as sequences having antimicrobial properties, and everything else as non-AMPs (negative sequences). To balance the set of positive and negative sequences, we selected a subset of the non-AMPs by clustering. The final version of *D*, used as our main dataset, contains 5922 AMPs and 5820 non-AMPs sequences. The plots in Fig. 3 show that in *D*, AMPs are shorter sequences than non-AMPs.

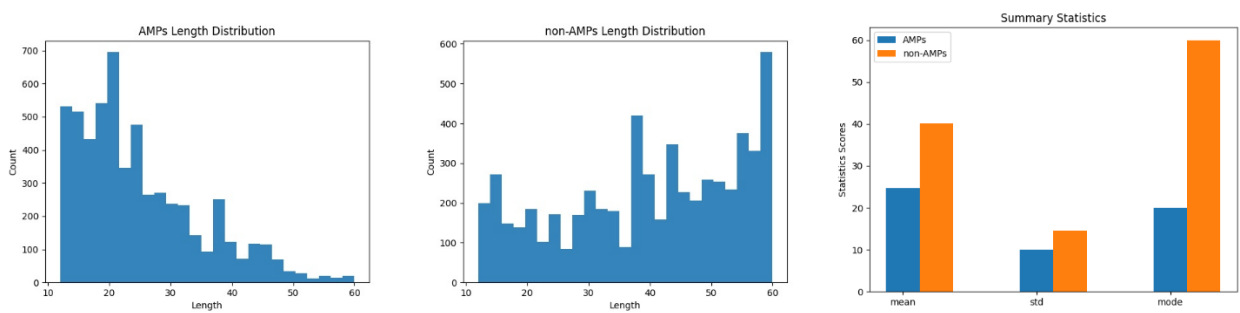

**Fig. 3. Summary Length Statistics over AMPs and non-AMPs Sequences. A) Length distribution of AMPs. B) length distribution of non-AMPs. C) Comparison between the two groups**

**Comparison between GraphCC and other selection baselines**

We implemented two GraphCC-based methods: *GraphCC Medoids* and *GraphCC Top-Scoring* (see details in Method section). To compare the effectiveness of these GraphCC-based methods, we established 4 additional selection method baselines: Random Selection, Select Algorithm, Sequences Medoids and Sequences Medoids Top-Scoring. GraphCC approaches and the baseline methods are described in detail in the Methods section. We present the results of GraphCC against other established baselines. For visualization purpose, we use the following abbreviations for the results of the different selection methods we described above: **RandomSeq** (*Random Selection*), **SelectSeq** (*Select Algorithm*), **SeqMedoids** (*Sequences Medoids*), **SeqMedoids TopScoringSeq** (*Sequences Medoids Top-Scoring*). **GraphCC MedoidsSeq**, and **GraphCC TopScoringSeq** stand respectively for *GraphCC Medoids*, and *GraphCC Top-Scoring*. The figures describe cumulative scores of best generated candidates selected up to the current round $r_t$.

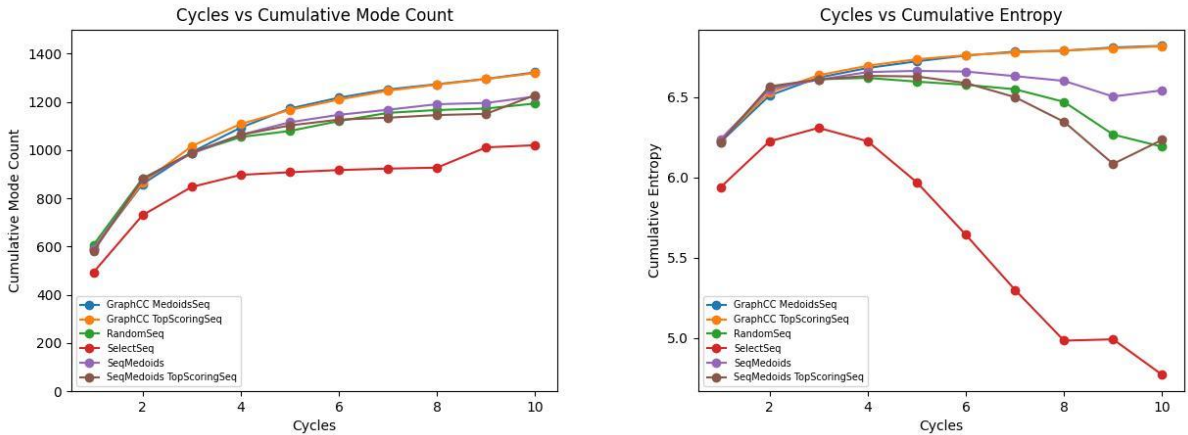

**Fig. 4. Performance on defined metrics. a) Cumulative Mode Count. b) Cumulative Entropy Scores**

We used the number of rediscovered modes and inter-modes entropy to measure cumulative performance of the generative models, in which higher number of rediscovered modes and higher inter-modes entropy indicate better performance. Fig. 4a and Fig. 4b report the curves of the mode count and the entropy values of different methods over the rounds. A higher score signifies a better performance, and the GraphCC-based methods are leading both the mode count and the entropy, in the least number of rounds while continuously improving.

We denote "*B*" the final sets of candidates, proposed by each selection method, obtained after the 10 rounds of active learning. To compare equal number of sequences, we define "*B'*": a subset of *B* generated by each method, randomly sampled, and containing the same number of AMPs as *D2*. After the cumulative performance, we computed global (Needleman-Wunsch) (see **Methods** section) and local (Smith-Waterman) (see **Methods** section) alignments between *B*, and our initial set *D2* (Fig. 5). These scores measure similarity between two sets of sequences, with more similar sets receiving higher scores. The GraphCC approaches generated lower scores than other methods, indicating that GraphCC indeed favors the sampling of more diverse AMP candidates that differ more from peptide sequences in our starting (initial) training dataset *D2*, and that it infers a higher diversity of candidates in the selection process.

Additionally, we compared the length distribution of *B'* and *D2* (Fig. 6) and computed the Kullback–Leibler divergence ($D_{KL}$) ((13), Table 1). $D_{KL}$ is a statistical *asymmetric* measure widely used to compare how different is the probability distribution *P* from another probability distribution *Q*. In our case, as we are working with discrete values, the $D_{KL}$ is defined as follows:

$$D_{KL}(P||Q) = \sum_{x \in X} P(x) log\left(\frac{P(x)}{Q(x)}\right),$$

where *X* is the relative probability from *Q* to *P*, *P* is the probability distribution over the lengths of *B'*, and *Q* is fixed to the lengths over the initial set *D2*. The lower the value of $D_{KL}$, the more similar are the distributions *P* and *Q*.

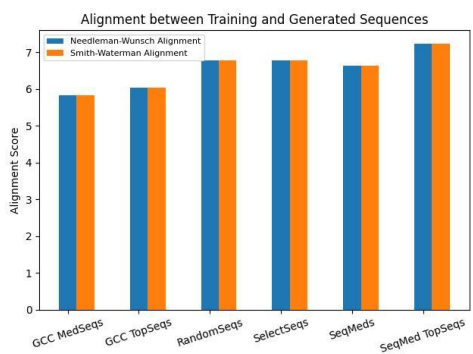 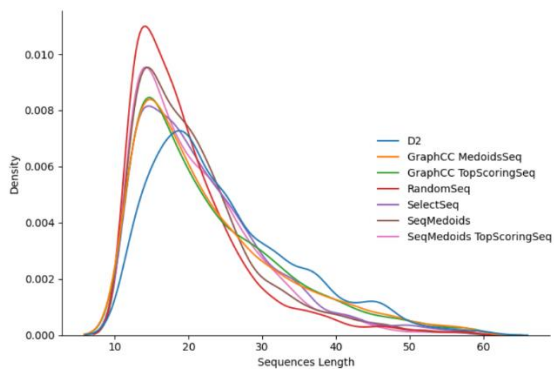

**Fig. 5. Alignments between *B* and *D2***      **Fig. 6. Length Distribution of *B'* and *D2*** 

| Selection Method | $D_{KL}(P|Q)$ |
|---|---|
| **GraphCC MedoidsSeq** | **0.2297** |
| **GraphCC TopScoringSeq** | **0.2336** |
| RandomSeq | 0.2337 |
| SelectSeq | 0.2371 |
| SeqMedoids | 0.2444 |
| SeqMedoids TopScoringSeq | 0.2411 |

**Table. 1. $D_{KL}$ between sequences length distribution of *B'* and *D2***

The results from both Table 1 and the Fig. 6 reveal that the length distribution of sequences generated by GraphCC approaches are the closest to the length distribution of sequences in the initial set *D2*. This means that GraphCC samples and selects sequences following the same distribution as *D2*, is consequently prompter to select novel and real candidates, while enhancing their generation. Moreover, using the DBAASP, the authors of ClaSS (10) generated a list of new AMPs. Among those sequences, the authors identified two sequences YLRLIRYMAKMI-CONH2 (YI12, 12 amino acids) and FPLTWLKWWKWKK-CONH2 (FK13, 13 amino acids), to be the best with the lowest Minimum Inhibitory Concentration (MIC) values. Neither of those two best-proposed sequences are in our initial set *D2* nor in the list of sequences that our GraphCC-based methods generated, which speaks of the novelty generation of our approach.

To confirm the efficiency of our GrapCC methods, and the diversity of generated solutions, we calculated several common biophysical and other sequence-based features for each peptide and their distributions between the methods using prediction algorithms as implemented in the ProtParam module of the Biopython library (14). Each GraphCC method generated 10,000 AMP sequences. For comparison analysis, we also considered the set of sequences generated by the next best performing method (SeqMedoids) which also generated 10,000 sequences and the sequences comprising the *D1* and *D2* sets. We considered the following peptide properties: (a) *aromaticity* which represents the proportion of amino acids that have aromatic sidechains in the whole peptide, (b) *charge at pH* (where we considered pH = 7) which represents the net charge on the peptide at pH of 7, (c) *flexibility* represents the amount of conformational flexibility of the peptide in solution, (d) *instability index*, presence of motifs can render a peptide more vulnerable to various proteolytical enzymes, (e) *isoelectric point* (pI) is the pH of a solution at which the overall charge on the protein is 0, (f) *molecular weight* measured in gram/mol representing the actual mass of the peptide, (g) *gravy* which is a property indicating the hydrophobicity (15) of the peptide, (h) *secondary structure function* which indicates the fraction of amino-acids in a alpha-Helix, beta-Sheet, and beta-Turn conformation and offers a measure for AMP conformational/structural diversity. We also computed the average percentage of each amino acid per peptide, to see if there are any shifts in amino acid distribution. Many known classes of AMPs show distinct combinations of these and other features (16). Of note, many of these predicted features are interrelated, for example a high aromaticity score likely also results in a high gravy score and charge of a peptide at pH 7 is also related to its pI. Here, we consider a wide distribution of these predicted features within a set as a coarse indication that the set will also contain peptides showing a broad distribution of structural and biological properties.

We conduct analyses to measure the properties listed above and to compare their respective results to those of our *D1 set* and the initial training set (*D2*). Our analyses and comparisons will be mainly with respect to *D1* set, as it is the set from which we designed set of peptides prototypes (see **Methods** section). Figure 8 presents the fraction of $\alpha-$Helix, $\beta-$Turn and $\beta-$Sheet (peptide secondary structures characteristics) in sequences generated by the GraphCC and SeqMedoids (the next best performing method) methods, compared to *D1* and the initial training set *D2*. Ideally, for these structural characteristics, we want our GraphCC methods to have their statistics as far from *D1* statistics as possible: this will imply the existence of structural differences between GraphCC sequences and the sequences of *D1*. Therefore, the diversity in structure characteristics will infer functional diversity. The main takeaway from Figure 8 is that GraphCC methods indeed offer more structural diversity; in terms of secondary structure (fraction of Helix, Turn and Sheet). In Figure 9 we show additional statistics of GraphCC, SeqMedoids, *D1* and initial *D2* sequences. Conversely to the context of the secondary structure, we want these additional statistics; for our GraphCC sequences to be as close to *D1* statistics as possible: this will mean that, statistically, the sequences generated by GraphCC have the desired biophysical properties as the sequences of *D1* and initial *D2*. Indeed, we can see that the distribution of the set of desired properties of GraphCC generated sequences, is mostly closer to the distribution of *D1*. We can then affirm that in general, our GraphCC methods statistically provide high quality, novel, and diverse candidates: the generated sequences have properties like our *D1* set, which allows us to say that these sequences could carry desirable antimicrobial properties and offer more structural diversity.

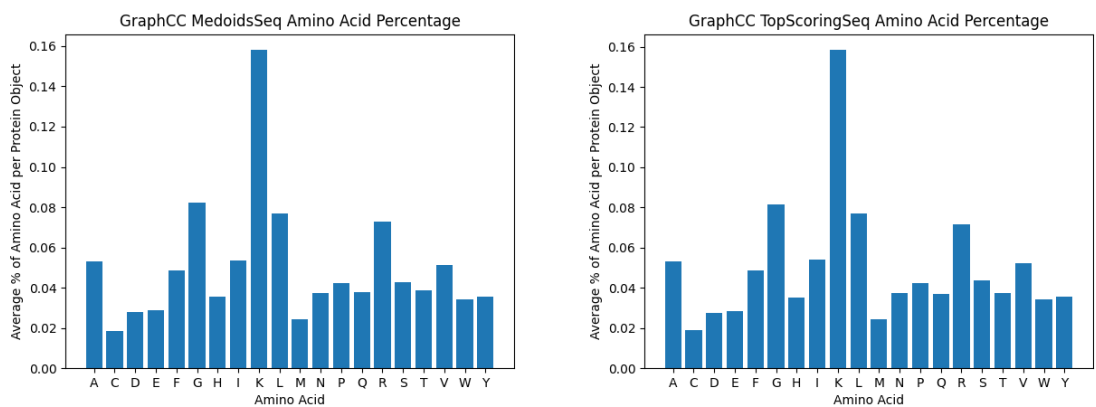

**Fig. 7. Average Percentage of each Amino Acid per peptide generated a) in GraphCC MedoidsSeq. b) in GraphCC TopScoringSeq**

The results in Fig. 7. shows a preference for lysine (K), justifiable by the fact that we have a similar trend in amino acid frequency in our initial dataset.

**Discussion**

In this study, we designed an active learning framework, with a graph connected component (GraphCC)-based selection method that enables the generation of novel and diverse peptide sequences.

Our results show that GraphCC performs better and offers more diversity than other baselines. Not only does it perform better, but it is also able to filter and select sequences that present structural diversity. We also show that it provides sequences which have biological properties closer to our *D1*.

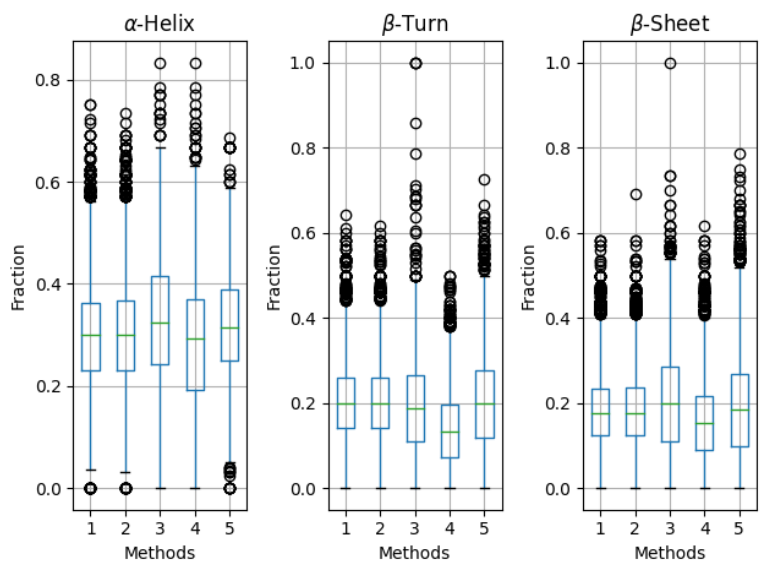

**Fig. 8. Percentage of Helix, Turn and Sheet of the sequences generated by GraphCC (1: GraphCC MedoidsSeq, 2: GraphCC TopScoringSeq) and SeqMedoids (5: SeqMedoids) methods, of the prototype and initial training sets (respectively 3: Initial (D1) Sequences, and 4: Initial (D2) Sequences).**

One limitation of GraphCC is that it tends to select short (12-20) length sequences, considered real even according to our initial dataset $D$. This favors a small number of connected components (2, 4, etc.) with huge sizes. Depending on the number of medoids selected per connected components, our dataset D2 will be very slowly augmented, which could impact the performance of the entire framework. *Algorithm 2* takes a batch of sequences, newly generated and creates a graph from it. In the given graph, two nodes (1 node = 1 AMP sequence) are linked only if the distance between them is less than a given distance threshold. If that criterion is satisfied, we can be sure that they belong to the same family/community in the graph. Applying the same logic to the whole graph yields non-overlapping communities, the connected components (CC). Each CC represents on its own a different type of sample, therefore, ideally, the higher the number of CCs, the more diversity. However, if the number of CCs is too large, this will mean that all or practically all samples are far from one another, and hence forming one CC on their own: that is a situation we do not want either. To remedy that, sampling a greater diversity of lengths (possibly middle) is desired. Peptides used in our experiments, made only of 20 natural amino acids have properties that make them not perfect drugs. However, our developed active learning framework performing well on natural amino acids can be easily expanded to include non-natural amino acids and other modular chemical building blocks to create peptidomimetics and macrocycles with improved drug-like properties and hence will help to create more useful drugs, faster and more efficiently.

**Materials and Methods**

**Active-learning Framework**

Our multi-round experiments are implemented through an active learning framework described in Fig. 1. $G$ is an LSTM (17) generative recurrent neural network, trained to generate one character (one amino acid) at the time, while $S$ is another LSTM classifier trained to differentiate between AMPs and non-AMPs (respectively labeled as 1 and 0). As we developed and compared many approaches, we then built a pair ($G$, $S$) for each of them, enabling us to better access their respective efficiencies. A recurrent neural network (RNN) is a neural network which works with sequential data (for instance data where positional or temporal information is crucial). LSTM uses its "memory" to integrate knowledge or information from past inputs to process the current input. As result, the output of an RNN depends on the prior elements within the sequence and takes order of the elements into consideration. This is particularly relevant in our setup because the sequence of amino-acids generated so far determines the next one amino-acid to be generated.

$D$ is split into two non-overlapping sets *D1* and *D2*. *D1* is used to create the wet-lab simulator that distinguishes AMPs from non-AMPs (hereafter called oracle *O*), and *D2* is used to train *G* and *S*. This is to ensure *G* and *S* *do not* see data used for O, enabling a fair judgment from the latter. *D1* and *D2* each contain 2961 AMPs, and 2909 non-AMPs sequences. At each round, the best sequences generated by *G* were selected and added into *D2*. Both *G* and *S* are retrained from scratch, at each round $r_t$ using *D2*. At each round $r_t$, we use the *G* to generate 10,000 sequences. We call that set $D_{new}$.

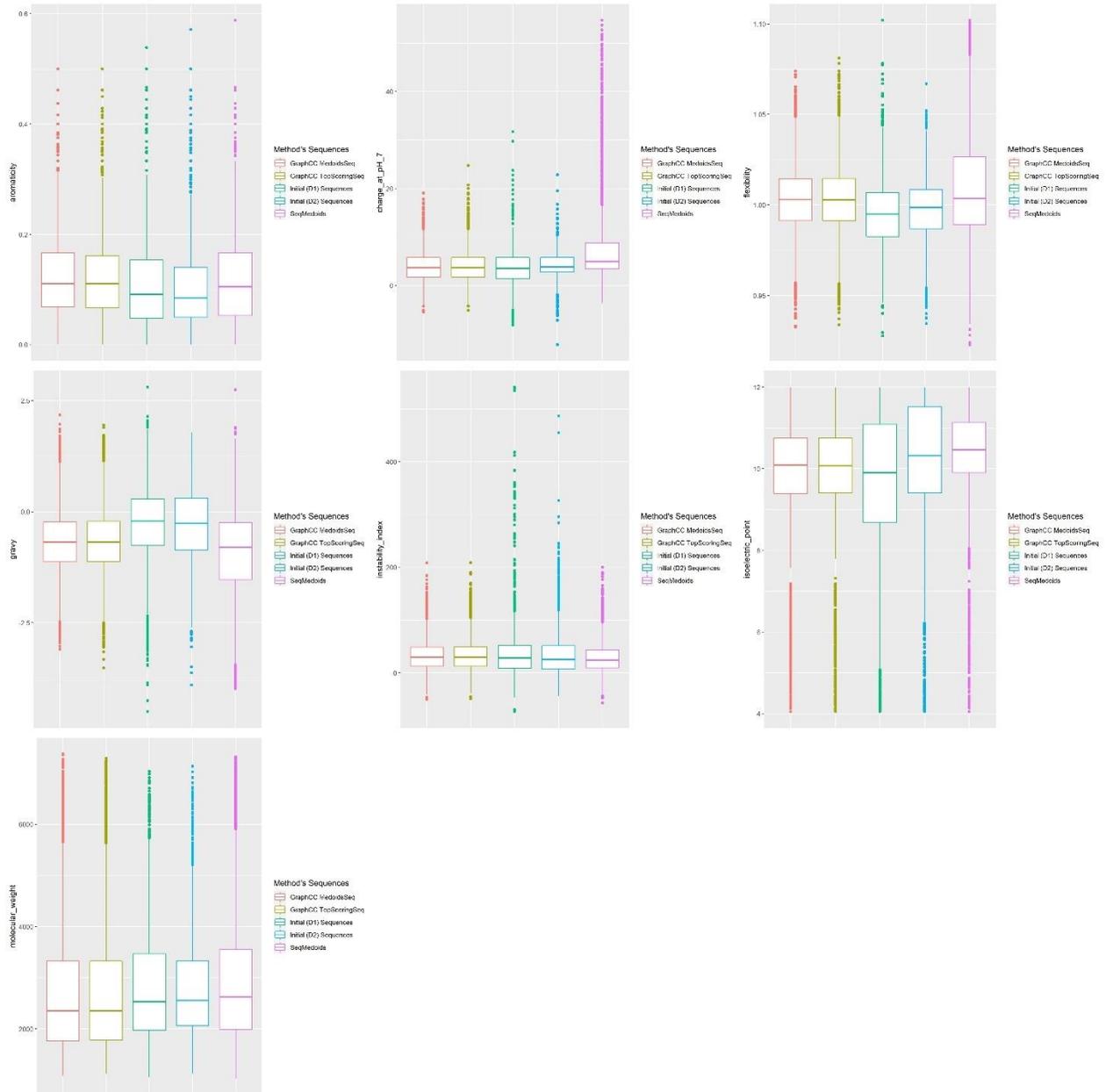

**Fig. 9. Properties of sequences generated by GraphCC and SeqMedoids methods. In green and light blue are the statistics of the prototypes set (D1) and initial training set (D2).**

We tried several approaches to select the best subsets of sequences from $D_{new}$: *Random Selection* (we select randomly *k* sequences from $D_{new}$), *Select Algorithm* described in Algorithm 1, *Sequences Medoids*, and *Sequences Medoids Top-Scoring*.

In *Sequences Medoids* method, $D_{new}$ are clustered into $k$ clusters using the k-medoids (18, 19) algorithm with a distance matrix $A$. $A$ contains the pairwise distance $d_{(s_1,s_2)}$ between all pairs of amino acid sequences in $D_{new}$. In our experiments, edit-distance (11) is used as distance metric $d$. After the clustering step, the medoids (or prototypes) of the $k$ clusters are picked. In the *Sequences Medoids Top-Scoring* method, after splitting $D_{new}$ into $k$-clusters using the k-medoids algorithm, instead of selecting the medoid of each cluster, the top-scoring sequence of each cluster was selected, using the score function $v(x)$ defined in Algorithm 1. This process selects the top-$k$ scoring sequences.

Alternatively, we tried 2 approaches using the GraphCC algorithm:

a) GraphCC Medoids: using Algorithm 2, from each connected component $CC_1, \ldots, CC_n \in G_{optimal}$, we select $k_i$ prototype sequences, $\sum_{i=1}^{n} k_i = k$. As the number of connected components vary, $k_i$ vary as well, and are dynamically computed using the size of the CC.
b) GraphCC Top-Scoring: from each connected component $CC_1, \ldots, CC_n \in G_{optimal}$, using the score function $v(x)$ defined in Algorithm 1, we select the top-$\delta_i$ sequences, $\sum_{i=1}^{n} \delta_i = k$. $\delta_i$ vary, and are dynamically computed using the size of the C

A high-level overview of the entire full-loop process is described in Algorithm 3.

**Generated Candidates Evaluation**

To construct $O$, we proceeded as follows:

a) we selected $l = 1500$ medoids from the set of AMPs in $D1$, using the $k$-medoids method using $d$. We denote $D_{prototypes}$, the set of the selected medoids, as the set of prototypes.
b) we defined then non-parametric reward function, $O(s)$ for multi-round experiments: $O(s) = \exp\left(-\min_{s' \in D_{hard}} d'_{s,s'}\right)$, with $d'_{s,s'} = \frac{d(s,s')}{len(s')}$

The main goal of $O$ is to rediscover the set of prototypes. In other words, $O$ compares on a sequence-level the generated sequences with the prototypes and serves as indicator of how *realistic* generated sequences are.

Let $M$ be the set all methods described above. At each round $r_t$, using $D_{new}$, we selected the best samples of each method. Let us call $b_m$, the set of best candidates selected by a method $m$. At each round, we selected 2000 samples using the random selection and the Select algorithm. However, for GraphCC and Sequences Medoids approaches, we could not define a fixed size. This was because: (1) in the GraphCC approaches the number of CCs and the size of each CC is not predictable beforehand; (2) with the medoids, even though we could fix the number of medoids, we cannot predict the size of each respective clusters. To make fair comparison (and equal the number of $O$ calls per method), we randomly sample $b$ sequences across all methods, where $b = min_{m \in M} size(b_m)$. For evaluation, and comparison across methods, we then select the top-1000 among the $b$ randomly sampled sequences (based on $O(s)$ scores) of each method. $O$, the wet-lab simulator, is the block that scores and labels those final best candidates.

**Diversity Measurement**

As our goal is to foster the discovery of new structures, it is very important for us, to also put the focus on structural ***diversity*** of the molecules. We care about diversity because a DL generative model would probably be useless for our purpose of discovering new AMPs, if it does not produce sequences that are relatively novel to known AMPs.

The diversity of **GraphCC** comes from the connected component methods on graph, where we first constructed a graph of AMPs based on structural similarity. One difference between **GraphCC** and existing clustering approaches in general, is the stricter condition for clusters separation. AMPs from different CCs are more different from one another (because two CCs in a graph are separate and disconnected from each other by some threshold of similarity). Conversely, other approaches only try to ensure that intra or inter-cluster similarity is big). Furthermore, in this study, we pick AMPs from each CC to ensure diversity of the proposed molecules.

We evaluate diversity in three ways:

a) **Metric 1 (Mode Count)**: the number of prototypes, to which, at least 1 candidate $\in B$ (the cumulative set of best AMPs sequences selected up to the current round $r$) is close to.
b) **Metric 2 (Mode Entropy)**: the entropy $H$ of the empirical distribution over prototypes:
   a. $\hat{p}(s_i) = \frac{1}{|B|} \sum_{s \in B} I\left[s_i = \operatorname{argmin}_{s' \in D_{prototypes}} d'_{s,s'}\right]$
   b. $H(\hat{p}) = -\sum_{s_i \in D_{prototypes}} \hat{p}(s_i) \log \hat{p}(s_i)$
c) **Metric 3**: Average global (Needleman-Wunsch (20)) and local (Smith-Waterman (21)) alignments scores between the training sequences of *D2* and the best candidates selected by each method we described in our paper. We used the Biopython (14) library to compute both alignment scores.

In our *O* and diversity metrics definitions, we normalized the distance of sequences to the prototype, by the length of the prototype. This was inspired from well-known alignment metrics like TM-align (22). For a higher diversity, we ideally want our methods to have:

> **Metric 1 (a mode count of $l$)**: this would tell us how many prototypes the methods tried to rediscover. For diversity, we do not want best selected sequences to be around few prototypes. Therefore, high values will not only indicate higher diversity, but also will tell us whether best sequences generated by the methods are novel and real enough.

> **Metric 2 (a high entropy $H$)** (close to $-\log \frac{1}{l}$): this is a direct consequence of a high mode count. In fact, the more spread are selected sequences over $D_{prototypes}$, the higher the entropy score will be.

> **Metric 3 (low pairwise alignment scores)**

**Algorithms of selection**

In this section, we describe the core algorithm of all selection methods we tried. We consider as baselines any method described above, that does not include the use of GraphCC.

---

**Algorithm 1** Select Algorithm

**Parameters**: $S, d, k, D_{new}, \gamma$: distance threshold
1: Scoring function $v(x) = \hat{S}(x) + \alpha \varepsilon(x)$, where $\hat{S}(x)$ is prediction for a given AMP sequence $x$, and $\varepsilon(x)$ its uncertainty, computed with MC-Dropout    //the reason behind introducing uncertainty in the process, is to encourage exploration more. Optimally, enhancing exploration could/should lead potentially to more diversity.
2: $D_{new} \leftarrow sort(D_{new})$ such that $\forall x_i, v(x_i) > v(x_{i+1})$
3: $O_{best} = \{x_1\}$ //we initialize our batch by the top-scoring sequence
4: $N = |D_{new}|$
5: **for** $i \leftarrow 2$ to N **do**:
6:     $\delta = \min_{x_j \in O_{best}} d(x_i, x_j)$
7:     **if** $\delta > \gamma$ **then**:
8:         $O_{best} \leftarrow O_{best} \cup \{x_i\}$
9:         **if** $|O_{best}| = k$ **then**:
10:             exit, and return $O_{best}$
11: **return** Select($S, d, k, D_{new}, \gamma/2$) /*If we fail to find $k$ sufficiently distant elements of $D_{new}$, we reduce $\gamma$ and retry.*/

**Algorithm. 1. Algorithm for the Select Method.** This method ensures that samples selected in the final batch are very distant from one another, hence very long. Everywhere the function $v(x)$ has been used, $\alpha = 0.3$. To measure uncertainty, we used MC-Dropout (23) i.e., we ran the selector $\hat{S}$, $n = 10$ times. As we used MC-Dropout, the dropout layer of the model is activated, introducing hence randomness, yielding in $n$ different results. We then computed $\epsilon(x)$ as the standard deviation of the $n$ scores.

---

**Algorithm 2** GraphCC Algorithm

**Parameters:** $D_{new}$, $d$, $R(s, s', t)$, $T$
- $R(s, s', t) = 1$ if $d(s, s') \leq t$, else $0$ // This generates a graph with nodes and edges. $R(s, s', t) = 0$ means absence of edge between $s$ and $s'$ and $1$ the opposite
- $T$: set of distance thresholds

**Output:**
- $O_{best}$: set of CC-medoids sequences

1: $G = \{G_i \mid G_i = R(s, s', t_i), t_i \in T\}$ //all possible graphs given different distance thresholds
2: $G_{optimal} = \{G' \text{ such that } |CCs(G')| = max_{G_i \subset G}|CCs(G_i)|\}$ //we speculate that the degree of diversity is proportional to the number of CCs, i.e the higher the number of CCs, the more diverse is the graph. Therefore, here we are selecting the optimal graph as the one who gives the highest number of CCs
3: **for** $cc \in CCs(G_{optimal})$ **do:**
4:     initial_medoids = $\{m_1, ..., m_k\}$ //set of the k sequences serving as initial prototypes for the current $cc$ - randomly chosen
5:     distance_matrix = $M_{n,n}$ where $n = |cc|$ such that $M_{ij} = d(s_i, s_j)$ with $s_i, s_j \in cc$ //distance of pairwise distances between sequences of the $cc$
6:     km = kmedoids(distance_matrix, initial_medoids) //following computations are made using the PyClustering Library (24)
7:     km.process()
8:     centers = km.get_medoids()
9:     $O_{best} \leftarrow O_{best} \cup \{\{x_i\} \in centers\}$
10: **return** $O_{best}$

---

**Algorithm. 2. Algorithm for the GraphCC Method.** This method takes a set of sequences newly generated and constructs an optimal graph. An optimal graph is the one bringing high and reasonable number of Connected Components, hence diversity. From each CC, we then select prototypes, with the conditions explained in methods description. At the end, we have the set of "most representative" candidates from each cluster, pulled together as best prototypes of the optimal graph created.

---

**Algorithm 3** Full-Loop (An entire round of training)

**Parameters:** $G$ trained on $D2$ augmented with previous rounds best selected sequences, $S$ trained on $D2$, Algorithm of selection $\hat{D}$ (any of the techniques described above), Hard-Oracle $O$.

1: Generate $D_{new}$ with $G$
2: $D_{best} \leftarrow \hat{D}(D_{new}, S, k', d, v, \gamma)$ (set of best candidates returned by $\hat{D}$)
3: $scores \leftarrow \{\}$
4: **for** $x_i$ in $D_{best}$ **do:**
5:     $g_{x_i} = O(x_i)$ (Oracle score for the sequence $x_i$)
6:     $scores \leftarrow scores \cup \{x_i, g_{x_i}\}$
7: select the top-$k'$ sequences based on the scores $g_{x_i}$.
8: $D2 \leftarrow D2 \cup \{\text{top-}k'\}$
9: retrain $G$ and $S$ on $D2$, and repeat steps 1-7

---

**Algorithm. 3. Algorithm for a full-Loop (an entire round of training).** The Algorithm 3 represents the full step for training one round of our setup. It works across all methods described with represent $\hat{D}$ in the algorithm. It selects their best candidates and appends it to their respective set D2 (respective to each task i.e., initially, at round $r_1$ all methods have the same D2. But from round $r_t, t \geq 2$, each method has its own D2, consisting of the initial D2, plus its best selected candidates from $r_1, ..., r_{t-1}$).

## Hyperparameters of Models

### Generative models
For the generative models, we used the following hyperparameters: *input_size*=22 (20 amino acids, SOS (Start of Sequence), and EOS (End of Sequence)), *hidden_size*=128, *output_size=input_size*, *n_layers* (number of layers) =1, *batch_size*=256, *lr* (learning_rate) =1e-5, *wd* (weight_decay) =1e-5. SOS is the prompt given to the generative models to start generating characters (each of them represent an amino acid), EOS tells the generative model when to stop the generation.

### Selector Models
For the selector (prediction) models, we used the following hyperparameters: *input_size*=22 (20 amino acids, SOS (Start of Sequence), and EOS (End of Sequence)), *hidden_size* = 64, *embedding_size*=64, *output_size*=2 (2 scores corresponding to the probabilities of the input sequence being AMP or not), *n_layers* (number of layers) = 1, *batch_size*=256, *lr* (learning_rate) = 1e-5, *wd* (weight_decay) = 1e-5.